\documentclass[aps,amsmath,amssymb,twocolumn,showpacs,prl,superscriptaddress]{revtex4}

\usepackage{graphicx}
\usepackage{dcolumn}
\usepackage{bm}
\usepackage{psfrag}

\begin{document}

\title{Multiphoton above threshold effects in strong-field fragmentation}

\author{C.~B.~Madsen}
\affiliation{Lundbeck Foundation Theoretical Center for Quantum
System Research, Department of Physics and Astronomy, Aarhus University, DK-8000 Aarhus C, Denmark}
\affiliation{J.~R.~Macdonald Laboratory, Kansas State University, Manhattan, KS, 66506-2604, USA
}

\author{F.~Anis}
\affiliation{J.~R.~Macdonald Laboratory, Kansas State University, Manhattan, KS, 66506-2604, USA
}

\author{L.~B.~Madsen}
\affiliation{Lundbeck Foundation Theoretical Center for Quantum
System Research, Department of Physics and Astronomy, Aarhus University, DK-8000 Aarhus C, Denmark}

\author{B.~D.~Esry}
\affiliation{J.~R.~Macdonald Laboratory, Kansas State University, Manhattan, KS, 66506-2604, USA
}

\date{\today}
\begin{abstract}
We present a study of multiphoton dissociative ionization from molecules. By solving the time-dependent Schr\"{o}dinger equation for H$_2^+$ and projecting the solution onto double continuum scattering states, we observe the correlated electron-nuclear ionization dynamics in detail. We show --- for the first time --- how multiphoton structure prevails as long as the energies of all fragments are accounted for. Our current work provides a new avenue to analyze strong-field fragmentation that leads to a deeper understanding of the correlated molecular dynamics.
\end{abstract}


\maketitle

Despite more than 20 years of scrutiny, strong-field dissociative ionization of molecules is still not completely understood. Understanding this process, however, would provide insight into how the energy deposited in the molecule by an intense laser pulse is shared between the electrons and the nuclei via their correlated motion. A large part of the challenge in investigating this process is that the dynamics of an ionized electron is not easily treated by the usual Born-Oppenheimer (BO) approximation. Complicating the issue further is the fact that the final state lies in at least a double continuum, likely Coulombic, comprised of the free nuclear and electronic motion which raises fundamental questions about the analysis.

Because ab initio calculations of dissociative ionization require going beyond the BO approximation, the vast majority of intense field dissociative ionization calculations have been carried out for the simplest molecule: H$_2^+$ (see, for example, Refs.~\cite{PhysRevA.52.2977,posthumus:rpp:2004,PhysRevA.80.023426}). Even for this system, however, full-dimensional solutions of the time-dependent Schr\"{o}dinger equation (TDSE) have not yet been obtained in an intense field as they pose an immense computational challenge. Consequently, the dissociative ionization calculations that have been done simplified the problem even further through ad hoc reductions of the dimensionality or other severe approximations.

Nevertheless, intense field processes identified via H$_2^+$ studies --- even reduced dimensionality ones --- are now understood to occur in other systems as well. Dissociation via bond softening and above threshold dissociation (ATD), for instance, were first identified in H$_2^+$ and have now been seen in other systems like CO$^{2+}$~\cite{PhysRevA.81.061401}, O$_2^+$~\cite{PhysRevA.83.053405}, Na$_2^+$~\cite{PhysRevA.78.021403}, and N$_2^+$~\cite{PhysRevA.78.033430}.

Ionization mechanisms such as charge-resonance enhanced ionization (CREI) have also been identified in H$_2^+$~\cite{PhysRevA.52.R2511,PhysRevA.57.1176} and applied to other systems~\cite{PhysRevA.76.053409,PhysRevLett.107.063201}. More recently, two new models have been proposed to explain unexpected structure measured in the nuclear kinetic energy release (KER) spectrum following ionization of H$_2^+$. In one model, the interference of two dissociation pathways leads to the modulation of the KER spectrum  via CREI~\cite{PhysRevLett.98.073003,PhysRevA.76.013405}. The other model, named above threshold Coulomb explosion (ATCE), is based on the Floquet potentials obtained by dressing not only the field-free BO potentials but also the $1/R$ Coulomb explosion curve with photons from the laser field (see Fig.~\ref{fig:fig3})~\cite{PhysRevLett.97.013003}.  The observed structure in the KER spectrum is thus due to absorption of different numbers of photons~\cite{PhysRevLett.97.013003,PhysRevA.82.043409}. The predictions of the models deviate for low intensities and there is no consensus on which is correct.  So, in addition to answering fundamental questions about electron-nuclear correlations, TDSE solutions --- along with an accurate way to analyze them --- are needed to resolve this controversy.

Key to understanding electron-nuclear correlations is the identification of an appropriate physical observable.  It has long been recognized~\cite{0034-4885-66-9-203} that a particularly useful observable for this purpose is the joint energy spectrum (JES).  For instance, experiments at the Advanced Light Source have shown how the simultaneous measurement of the KER and the energy of a freed electron can be used to extract details about the single-photon-induced breakup of molecules such as C$_2$H$_2$~\cite{0953-4075-41-9-091001} and CO~\cite{PhysRevA.81.011402}. Joint energy spectra are also useful beyond electron-nuclear dynamics.  In fact, they provide insight whenever the final state involves three or more fragments.  An application to the double ionization of He in an intense laser field showed, for example, that the joint electron energy spectrum would reveal a clear separation of sequential from non-sequential processes~\cite{KTTaylor} and could thus help answer an outstanding question~\cite{PhysRev.164.1} in intense field physics.  While not exactly a JES, the goal of the well-known Dalitz plots~\cite{doi:10.1080/14786441008520365} is very similar.  Yet, even with so much evidence of their utility, JES following intense field dissociative ionization have not been studied.

Our goal in this Letter is to show that the JES does indeed provide considerable insight into strong-field dissociative ionization.  First and foremost, it shows clear multiphoton structure reminiscent of --- but distinct from --- above threshold ionization (ATI)~\cite{PhysRevLett.42.1127}, ATD~\cite{PhysRevLett.64.1883}, and ATCE that was not previously seen or anticipated.  But, because the JES  clearly shows how the energy is shared, it can also help answer the outstanding question of how much energy is taken by the electron during strong-field ionization. Finally, the JES may well provide an alternative and detailed avenue to time-dependent imaging of molecular processes.

While our conclusions generalize to any molecule, we will illustrate them here for dissociative ionization of H$_2^+$.  The simplicity of this system removes the theoretical approximations that would otherwise be present for multielectron or polyatomic species, providing an unambiguous demonstration of our ideas.  For this same reason, we will
also consider a reduced-dimension model of H$_2^+$ to allow for an essentially exact numerical solution of the TDSE.  Since our results are based on energy distributions and depend primarily only on energy conservation, they are not sensitive to the dimensionality, making the reduced-dimension H$_2^+$ the most transparent example for our purposes.

Specifically, we consider a linearly polarized laser field with the nuclei aligned along the polarization axis, including just the internuclear separation $R$ and the electronic coordinate $x$ in the direction of the laser polarization, measured with respect to the nuclear center-of-mass. 
The length gauge Schr\"{o}dinger equation then reads (atomic units are used throughout),
\begin{equation}\label{eq:TDSE}
i\frac{\partial}{\partial t}\Psi(R,x,t)=\left[H_\text{N}+H_e+\mathcal{E}(t)x\right]\Psi(R,x,t),
\end{equation}
where
\begin{align}
H_\text{N}&=-\frac{1}{m_\text{p}}\frac{\partial^2}{\partial R^2}+\frac{1}{R},& \\
H_e&=-\frac{1}{2}\frac{\partial^2}{\partial x^2}-\frac{1}{\sqrt{x_A^2+a^2(R)}}-\frac{1}{\sqrt{x_B^2+a^2(R)}}
\label{eq:softcore}
\end{align}
with $x_{A,B}=x\pm R/2$.
We have softened the Coulomb singularity with the function $a(R)$, chosen to reproduce the full-dimensional $1\sigma_g$ BO potential --- and thus the ionization potential~\cite{PhysRevA.53.2562,PhysRevA.67.043405,PhysRevLett.98.253003}. 
We use a laser electric field of the form
\begin{equation}
\mathcal{E}(t)=\mathcal{E}_0\sin^2(\pi t/\tau)\cos(\omega t),\quad 0\leq t\leq \tau\label{eq:sin2env}
\end{equation}
with $\tau=N(2\pi/\omega)$ in which $N$ is the number of cycles ($\tau_\text{FWHM}\simeq0.364\tau$). Results in this paper use $N=10$.

We solve Eq.~\eqref{eq:TDSE} using the finite element discrete variable representation~\cite{PhysRevA.63.022711,PhysRevE.73.036708,0953-4075-37-17-R01,Boyd99chebyshevand}.   The $(R,x)$ plane is divided into inner, $\vert x\vert\leq$ 50~a.u. and $R\leq$ 15~a.u., and outer, $50$~a.u. $\leq\vert x\vert\leq$ 1500~a.u. and 15~a.u. $\leq R\leq$ 50~a.u., regions.  Within each region, $R$ and $x$ each have a uniform element distribution with the element size based on the minimum expected de~Broglie wavelength.  The abrupt change in element size at the boundary between regions is eliminated by repeated application of three-point averaging. 
We propagate the solution using the short-time evolution operator,
\begin{align}
\Psi(R,x,t+\delta)&\approx e^{-i[H_\text{N}+H_e+\mathcal{E}(t+\delta/2)x]\delta}\Psi(R,x,t),
\label{eq:Uexp}
\end{align}
evaluated with the Lanczos algorithm~\cite{kuleff:044111}. This combination of techniques produces very accurate solutions across many orders of magnitude and, based on testing, gives an accuracy of at least two significant digits.

A critical aspect of calculating the electron-nuclear energy spectrum is the careful separation of the $p+p+e^-$ double continuum from the $p+H$ single continuum. We accomplish this separation by calculating the double continuum wavefunction $\psi^{\sigma_x}_{E_\text{N},E_e}$ within the BO approximation. This approach guarantees their orthogonality to the bound electronic states, and it incorporates the Coulomb interactions.
The probability of observing a KER of $E_\text{N}$ {\em and} an electron energy of $E_e$ is thus 
\begin{equation}\label{eq:Espectrum}
\frac{\partial^2 P}{\partial E_\text{N}\partial E_e}=\sum_{\sigma_x=g,u}\vert \langle \psi^{\sigma_x}_{E_\text{N},E_e}\vert \Psi(\tau)\rangle\vert^2,
\end{equation} 
where $g,u$ denotes gerade and ungerade symmetry, respectively. Explicitly, our approximation to the double continuum scattering states is
\begin{equation}
\psi^{\sigma_x}_{E_\text{N},E_e}(R,x)=\chi_{E_\text{N}}(R)\phi^{\sigma_x}_{E_e}(R;x)
\end{equation}
with $\chi_{E_\text{N}}$ and $\phi^{\sigma_x}_{E_e}$ energy normalized scattering states satisfying
\begin{align}
H_\text{N}\chi_{E_\text{N}}(R)&=E_\text{N}\chi_{E_\text{N}}(R),\label{eq:Enuc}\\
H_e \phi^{\sigma_x}_{E_e}(R;x)&=E_e\phi^{\sigma_x}_{E_e}(R;x)\label{eq:Eelec}
\end{align}
which are each solved using a variational  R-matrix formulation~\cite{RevModPhys.68.1015}.

Since this analysis method has not been applied before to strong-field processes, we emphasize its attractive features beyond the rigorous separation of the double continuum it provides.  Because it is based on using the energy eigenstates, the spectrum can be calculated as soon as the pulse is negligible.  So, while the resulting spectra are consistent with existing approaches such as a simple Fourier transform~\cite{PhysRevA.76.063407} and the scaled coordinate approach~\cite{PhysRevA.71.013411}, our  method is much cheaper computationally as we need not propagate the solution to macroscopic times --- the latter being necessary both to make the Fourier transform a good approximation for a Coulombic system~\cite{PhysRevA.76.063407} and to improve the approximate identification of the double continuum based on spatial position.

We solved Eq.~\eqref{eq:TDSE} for wavelengths from 230 to 650~nm, for intensities from 10$^{13}$ to 10$^{14}$~W/cm$^2$, and for a number of initial vibrational states. As measured by the Keldysh parameter~\cite{keldysh}, all cases are weakly within the multiphoton regime: neglecting nuclear motion and using an ionization potential of 1.1~a.u., the Keldysh parameter lies in the range 2--8.

\begin{figure}
\centering
\includegraphics[width=\columnwidth]{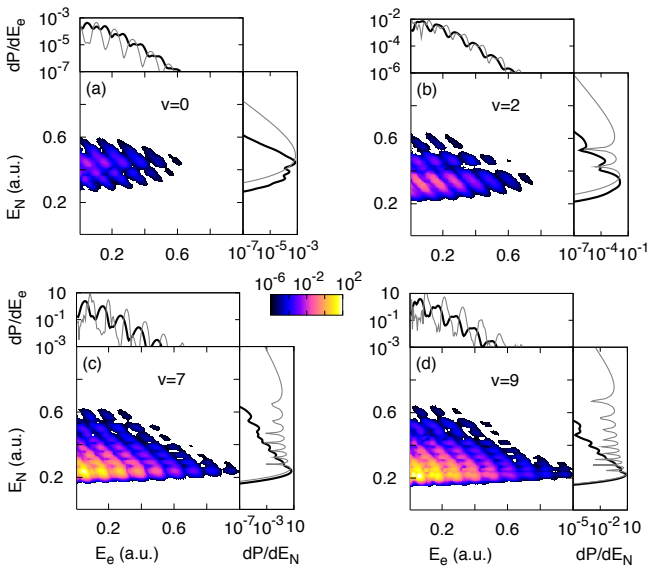}
\caption{(Color online) JES from Eq.~\eqref{eq:Espectrum} for H$_2^+$ exposed to a 400 nm, 8.8$\times10^{13}$ W/cm$^2$ laser pulse starting from (a) $v=0$, (b) $v=2$, (c) $v=7$, and (d) $v=9$.  The thick black curves in the top and side panels show the ATI and KER spectra, respectively, from integration of the JES. For comparison, frozen nuclei and reflection results (see text) are shown with thin grey lines.
}
\label{fig:fig10}
\end{figure}

Figure~\ref{fig:fig10} shows JES for H$_2^+$ exposed to a 400~nm laser pulse, starting from $v$=0, 2, 7, and 9 of H$_2^+$($1g$). 

\begin{figure}
\centering
\includegraphics[width=\columnwidth]{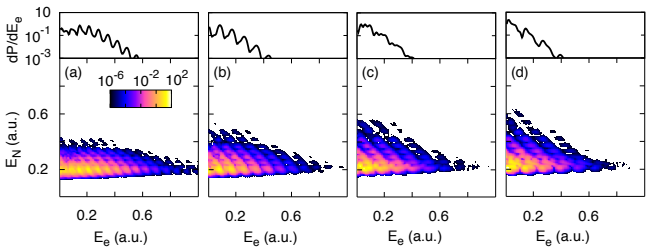}
\caption{(Color online) The JES from Eq.~\eqref{eq:Espectrum} starting from $v=9$ at $5.6\times10^{13}$ W/cm$^2$ as a function of  wavelength: (a) 650 nm,  (b) 506 nm,  (c) 400 nm, and (d)  363 nm.
}
\label{fig:figFreq}
\end{figure}

A characteristic feature of these density plots is the maxima along lines given by $E_v+n\hbar\omega=E_\text{N}+E_e+U_p$. Here, $U_p$ is the ponderomotive energy of the electron due to the presence of the laser field~\footnote{See Supplemental Material at ?URL?\label{footnote}}; and $E_v$, the energy of the initial state relative to the $p+p+e^-$ threshold. In other words, these plots are evidence of multiphoton absorption in the molecular double continuum. This conclusion has been confirmed via calculations at other wavelengths (see Fig.~\ref{fig:figFreq}).

Interestingly, these JES reveal much more than multiphoton features. For instance, in the present case they reflect strong electron-nuclear correlation. To understand this, we first note that --- as with any homonuclear diatomic molecule --- the laser field couples only to the relative motion of the electron and the nuclear center of mass. Thus, energy can be transferred into the relative nuclear motion only through interactions with the electron. Yet, for a given number of photons absorbed, there is a nonzero probability that the relative motion of the nuclei takes most of the energy, leaving the electron with almost no asymptotic kinetic energy. This behavior is revealed in Fig.~\ref{fig:fig10} by the broad distribution in $E_\text{N}$ at $E_e\sim0$ which indicates efficient energy transfer from the electron to the nuclei. This strong correlation could be important in other molecules, thus raising questions about neglecting nuclear motion even for short laser pulses.

In order to understand the additional structure apparent in Figs.~\ref{fig:fig10} and \ref{fig:figFreq}, we consider the strong-field approximation (SFA)~\cite{PhysRevA.22.1786} for the case of dissociative ionization~\cite{PhysRevA.76.033414} and approximate the final state of the nuclei with a plane wave. We note, in passing, that a somewhat similar extension of the SFA has been made to study the correlated dynamics of multi-electron atoms~\cite{doi:10.1080/09500340.2010.543958}.  For the present case, one can show that within the BO approximation, the SFA gives~[47]
\begin{equation}\label{eq:SFA}
\frac{\partial^2P_{\text{SFA}}}{\partial E_{\text{N}}\partial E_e}\propto\vert\tilde{\chi}_v(\sqrt{m_\text{p}E_\text{N}})\vert^2\vert\tilde{\phi}_0(\sqrt{2E_e})\vert^2F^{\mathcal{E}}_{E_\text{N},E_e},
\end{equation}
where $\tilde{\chi}_v$ and $\tilde{\phi}_0$ are the Fourier transforms of the initial vibrational and electronic wave functions, respectively. The presence of $\vert\tilde{\chi}_v\vert^2$ explains the origin of the features at constant $E_\text{N}$. The factor $\vert\tilde{\phi}_0\vert^2$ similarly shows that features at constant $E_e$ are also possible. Together, these factors show that the JES can be used to image the total electron-nuclear initial state. The factor $F^{\mathcal{E}}_{E_\text{N},E_e}$ in Eq.~\eqref{eq:SFA} contains all of the laser parameter dependence and thus accounts for the multiphoton lines in the JES.

In all previous studies of this system (see, for example, Refs.~\cite{PhysRevA.57.1176,0953-4075-45-8-085103,PhysRevA.82.043409,PhysRevLett.95.073002,PhysRevLett.98.073003}) --- and multiphoton dissociative ionization more generally --- it was the nuclear KER spectrum that was presented and not the JES.  Moreover, a substantial fraction of experimental and theoretical work still relies on the simple reflection method either to produce the KER spectrum from a nuclear wave packet upon ionization or to deduce the nuclear $R$-distribution from a measured KER spectrum~\cite{PhysRevLett.82.3416,PhysRevA.65.023403,0953-4075-40-3-002,PhysRevLett.108.073202}. Since the pulse lengths we consider are all short compared to the H$_2^+$ vibrational periods, the reflection method amounts to the mapping $d P/d E_N\propto\vert\chi_v(1/E_N)\vert^2/E_\text{N}^2$. 

With our accurate analysis of the double continuum, we can test this simple method for our model H$_2^+$.  Our KER spectra, obtained by integrating the joint spectra  over $E_e$, are thus shown in the side panels of Fig.~\ref{fig:fig10} along with the reflection method spectra. Although the reflection method accounts for some of the gross features of the KER, key features are not reproduced.  This failure suggests the violation of at least one of its foundational assumptions: (i) that the ionized electron is fast and leaves behind two bare protons that subsequently Coulomb explode,  (ii) that the nuclei have zero kinetic energy to start this explosion, and (iii) that the ionization rate is independent of $R$. 

Just as the KER spectrum has been used to study nuclear dynamics, much has been learned about electronic dynamics by looking at the electron's ATI spectrum. The existence of the ATI peaks, for instance, reflects the periodic launch of electron wave packets at each half-cycle of the laser pulse. Additional features in the spectrum arise from electronic structure and intra-cycle dynamics~\cite{PhysRevLett.108.193004}. The top panels of Figs.~\ref{fig:fig10} and \ref{fig:figFreq} show the results of integrating the JES over $E_\text{N}$, and, interestingly, the ATI peaks survive. Their survival is a consequence of the fact that the JES peak sharply at constant $E_\text{N}$.

For comparison, we have calculated the ATI spectra by solving the TDSE for our model H$_2^+$ with the nuclei at a fixed distance corresponding to the maximum of $\vert\chi_v(R)\vert^2$. The results are shown in the top panels of Fig.~\ref{fig:fig10} as well. We note that these ATI peaks are both more pronounced and shifted compared to those obtained from the full calculation, underscoring the importance of nuclear motion even for very short laser pulses.

From Figs.~\ref{fig:fig10} and \ref{fig:figFreq}, it is clear that the KER and ATI spectra --- being projections --- obscure much of the information visible in the underlying JES.  We thus suggest two different and complementary spectra for studying the dynamics: the total energy and energy sharing spectra.  

\begin{figure}
\begin{center}
\includegraphics[width=1.0\columnwidth]{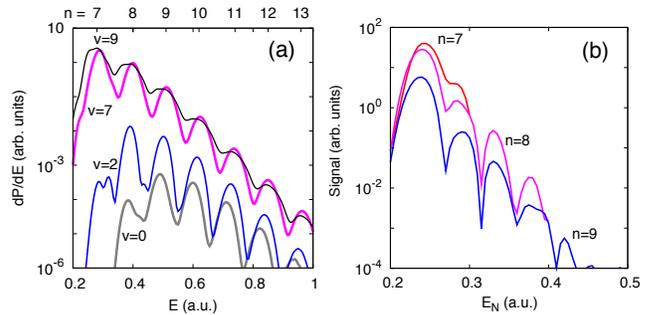}
\caption{(Color online) (a) The total energy ($E=E_\text{N}+E_e$) spectra for the JES in Fig.~\ref{fig:fig10}. (b) Cuts of the JES from Fig~\ref{fig:fig10}(c) demonstrate the energy sharing between the electron and nuclei starting
from $v=7$ for 7- ($E=0.3$ a.u.), 8- ($E=0.4$ a.u.), and 9-photon ($E=0.5$ a.u.) absorption.}
\label{fig:totalEnergy}
\end{center}
\end{figure}

\begin{figure}
\begin{center}
\includegraphics[width=0.8\columnwidth]{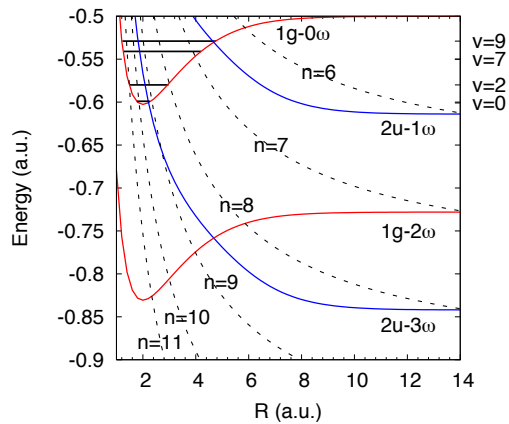}
\caption{(Color online) Dressed bound and ionization threshold potentials at 400 nm  for our 1D H$_2^+$ model. The horizontal lines indicate the energies of the states $v=0,2,7$, and $9$.} 
\label{fig:fig3}
\end{center}
\end{figure}

Since the multiphoton peaks occur along energy conservation lines, it is natural to integrate along lines of constant energy to obtain the total energy spectrum. Carrying out the integration for the JES of Fig.~\ref{fig:fig10}, we obtain the results shown in Fig.~\ref{fig:totalEnergy}(a). The opening of lower-$n$ photon channels with increasing $v$ apparent in the figure is readily predicted from the ionization-supplemented Floquet curves in Fig.~\ref{fig:fig3}.  In particular, the lowest peak visible for a given $v$ is determined from Fig.~\ref{fig:fig3} by the lowest-$n$ ionization curve that crosses the $1g-0\omega$ curve at a distance where $|\chi_v(R)|^2$ is non-negligible.  For instance, Fig.~\ref{fig:fig3} shows that $n$=6 crosses $1g-0\omega$, but it does so where $|\chi_v(R)|^2$ is negligible for all $v$ in Fig.~\ref{fig:totalEnergy}.  Consequently, the lowest peaks visible have $n$=7.  But, for $v$=0 and 2, even the $n$=7 crossing is not accessible, so their lowest $n$ is 9 and 8, respectively.  This finding lends support the validity of the ATCE model which is based on such a figure~\cite{PhysRevLett.97.013003,PhysRevA.82.043409}.

It is important to emphasize that we expect a figure like Fig.~\ref{fig:totalEnergy}(a) for {\em any} system so long as the energies for all the fragmentation degrees of freedom are included in the total energy.  This expectation follows from the fact that the occurrence of multiphoton features depends essentially only on energy conservation.  By extension, we thus expect to see multiphoton features like those in Figs.~\ref{fig:fig10} and \ref{fig:figFreq} in the JES for {\em any} system when all the fragmentation degrees of freedom are accounted for.  Moreover, the same conclusion is reached by generalizing the SFA result in Eq.~\eqref{eq:SFA} to other systems.  

These multiphoton features thus provide a natural binning, suggesting that we study the energy sharing spectrum for each photon peak.
Figure~\ref{fig:totalEnergy}(b) shows precisely this for the $v$=7 state.  Note that the structure from the vibrational state is still clearly visible. Since the spectrum for each $n$ peaks at the same $E_\text{N}$, the nuclei tend to take the same amount of energy --- the minimum predicted from Fig.~\ref{fig:fig3} --- while the electron tends to take all energy from the excess photons, explaining the survival of the peaks in the ATI spectrum. 

The tendency we note for the electron to take most of the excess photon energy may generalize to other systems, but further investigation is needed. What is clear, is that a straightforward extension of the SFA result in Eq.~\eqref{eq:SFA} will apply to other systems implying that the Fourier transform of the initial state will be imprinted on each energy sharing peak.

In this Letter, we have highlighted just some of the insights possible from the JES.  By illustrating them with a physically transparent reduced-dimension H$_2^+$ model, we have obtained some initial answers to several questions for dissociative ionization such as how the energy is shared between the electron and nuclei and how the JES might be used for imaging the total --- electron plus nuclear --- initial state.  Further study is certainly in order, and this Letter suggests many possible directions.  For instance, while our SFA extension already shows the imaging possibilities, extending it a little further hints that the JES might also be a sensitive indicator of the breakdown of the BO approximation in the initial state.  In addition, we believe that studying the JES for IR-pump and XUV-probe schemes will provide the additional data needed to establish the validity of the ATCE. As it happens, strong-field experimentalists are currently tantalizingly close to being able to measure three particles in coincidence with sufficient statistics to produce a JES~\cite{PhysRevLett.107.143004,PhysRevLett.108.073202}.

The authors thank J.~V.~Hern\'{a}ndez, J.~J.~Hua, H.~Lamm and H.~A.~Leth for fruitful discussions. This work was supported by the Danish Research Council (Grant No. 10-085430), an ERC-StG (Project No. 277767 - TDMET) and the Chemical Sciences, Geosciences, and Biosciences Division, Office of Basic Energy Sciences, Office of Science, U.S. Department of Energy.

\end{document}


\title{Supplementary material for the manuscript "Multiphoton above threshold effects in strong-field fragmentation"}

\author{C.~B.~Madsen}
\affiliation{Lundbeck Foundation Theoretical Center for Quantum
System Research, Department of Physics and Astronomy, Aarhus University, DK-8000 Aarhus C, Denmark}
\affiliation{J.~R.~Macdonald Laboratory, Kansas State University, Manhattan, KS, 66506-2604, USA
}

\author{F.~Anis}
\affiliation{J.~R.~Macdonald Laboratory, Kansas State University, Manhattan, KS, 66506-2604, USA
}

\author{L.~B.~Madsen}
\affiliation{Lundbeck Foundation Theoretical Center for Quantum
System Research, Department of Physics and Astronomy, Aarhus University, DK-8000 Aarhus C, Denmark}

\author{B.~D.~Esry}
\affiliation{J.~R.~Macdonald Laboratory, Kansas State University, Manhattan, KS, 66506-2604, USA
}

\maketitle

\section{Strong-Field approximation}
In this supplementary material, we discuss the extension of the SFA to the case of dissociative ionization. Our approach closely follows the methodology of~[37], but we approximate the final state of the nuclei with a plane wave [see Eq.~\eqref{eq:nuclei} below] instead of a Coulomb wave. We do this to consistently treat the electron and nuclei at the same level, since our TDSE calculations indicate that the effects of the Coulomb potential on the nuclei and the electron are equally important. More importantly, doing so allows us to derive the simple formula given in Eq.~(10) of the main text. This simple formula for the JES is consistent with the TDSE result. Further, it highlights the imaging application of the JES and potentially provides a much less computationally demanding means of computing the JES.

Within the S-matrix formalism, the joint energy distribution can be written as
\begin{equation}\label{eq:SFA}
\frac{\partial^2P_{\text{SFA}}}{\partial E_{\text{N}}\partial E_e}\propto\sum_{k_\text{N},k_e}\frac{1}{\vert k_\text{N}\vert\vert k_e\vert}\left\vert\int\, dt\langle\psi_f(k_\text{N},k_e,t)\vert V_\text{Ext}\vert\psi_i(t)\rangle\right\vert^2,
\end{equation}
where the sum is over four terms, corresponding to the positive and negative values of the final momenta $k_\text{N}$ and $k_e$ leading to the same energy. In each term, $V_\text{Ext}$ represents the laser molecule interaction, i.e., $E(t)x$ in the length gauge and $A(t)\partial/\partial x+A^2(t)/2$ in the velocity gauge, with $A(t)=-\int^tdt'\mathcal{E}(t')$ the vector potential. 

As the initial state, we use
\begin{equation}
\psi_i(R,x,t)=\chi_v(R)\phi_0(x)e^{-iE_vt},
\end{equation}
which presupposes the validity of the Born-Oppenheimer (BO) approximation and that the electronic wave function may be approximated near the nuclear equilibrium position as an $R$-independent function (note that improving this assumption by using linear combination of atomic orbitals will lead to the same result within the order of $m_\text{p}^{-1/2}$). These assumptions are justified in this work by the fact that all considered laser pulses have a duration short compared to the vibrational period. Consistent with the SFA, we employ a product of a plane wave (PW) and a Volkov wave for the final state
\begin{equation}
\psi_f(k_\text{N},k_e;R,x,t)=\chi_{k_\text{N}}^\text{PW}(R,t)\psi_{k_e}^\text{Volkov}(x,t).
\end{equation}
We proceed using the velocity gauge, where
\begin{align}
\psi_{k_e}^\text{Volkov}(x,t)&=\frac{1}{\sqrt{2\pi}}e^{ik_ex-i\frac{1}{2}\int^t(k_e+A(t'))^2dt'}\label{eq:nuclei}\\
\chi_{k_\text{N}}^\text{PW}(R,t)&=\frac{1}{\sqrt{2\pi}}e^{ik_{\text{N}}R-i\frac{1}{m_\text{p}} k_{\text{N}}^2t}.
\end{align}
Inserting these into Eq.~\eqref{eq:SFA}, we have
\begin{align}
\frac{\partial^2P_{\text{SFA}}}{\partial E_{\text{N}}\partial E_e}&\propto\sum_{k_\text{N},k_e}\frac{1}{\vert k_\text{N}\vert \vert k_e\vert }\vert\tilde{\chi}_v(k_\text{N})\vert^2\vert\tilde{\phi}_0(k_e)\vert^2
\nonumber\\ &\times\bigg\vert\int e^{i\left[(E_e+E_\text{N}-E_v)t+\int^t[A^2(t')/2+k_eA(t')]dt'\right]}\left[A(t)k_e+\frac{A^2(t)}{2}\right]dt\bigg\vert^2
\end{align}
with $E_e=k_e^2/2$, $E_\text{N}=k_\text{N}^2/m_\text{p}$, and $\tilde{\chi}_v,\tilde{\phi}_0$ denoting the Fourier transform of the respective wave functions. 

We note that in the length gauge, the Fourier transform of the electronic state is evaluated at momenta shifted by the vector potential at each time and thus does not separate outside the integral. Additionally, the velocity gauge interaction is replaced by that of length gauge. However, by evaluating the time integral using the saddle-point approximation, one will arrive at the same qualitative statements as we derive below using the velocity gauge.

Our current numerical investigations are in the regime where multiphoton structure is displayed. We may thus assume that the vector potential has the simple form
\begin{equation}
A(t)=A_0\cos(\omega t).\label{eq:cwField}
\end{equation}
We may then derive an explicit expression for the joint energy distribution:
\small
\begin{align}
\frac{\partial^2P_{\text{SFA}}}{\partial E_{\text{N}}\partial E_e}&\propto \sum_{k_\text{N},k_e}\frac{1}{\vert k_\text{N}\vert \vert k_e\vert}\vert\tilde{\chi}_v(k_\text{N})\vert^2\vert\tilde{\phi}_0(k_e)\vert^2\nonumber\\
&\qquad\times\bigg\vert\sum_nJ_n\left(\frac{k_eA_0}{\omega},\frac{A_0^2}{4\omega}\right)\int e^{i\left[(E_e+E_\text{N}+A_0^2/4-E_v+n\omega)t\right]}\left[A(t)k_e+\frac{A^2(t)}{2}\right]dt\bigg\vert^2\nonumber\\
&=\frac{4}{\sqrt{2m_\text{p}E_\text{N}E_e}}\vert\tilde{\chi}_v(\sqrt{m_\text{p}E_\text{N}})\vert^2\vert\tilde{\phi}_0(\sqrt{2E_e})\vert^2\nonumber\\
&\qquad\times\bigg\vert\sum_n(-1)^nJ_n\left(\frac{\sqrt{2E_e}A_0}{\omega},-\frac{A_0^2}{4\omega}\right)\int e^{i\left[(E_e+E_\text{N}+A_0^2/4-E_v-n\omega)t\right]}\left[A(t)\sqrt{2E_e}+\frac{A^2(t)}{2}\right]dt\bigg\vert^2\nonumber\\
&=\vert\tilde{\chi}_v(\sqrt{m_\text{p}E_\text{N}})\vert^2\vert\tilde{\phi}_0(\sqrt{2E_e})\vert^2F^{\mathcal{E}}_{E_\text{N},E_e},\label{eq:SFACW}
\end{align}
\normalsize
where
\small
\begin{align}
F^{\mathcal{E}}_{E_\text{N},E_e}&\equiv\frac{4}{\sqrt{2m_\text{p}E_\text{N}E_e}}\nonumber\\
&\qquad\times\bigg\vert\sum_n(-1)^nJ_n\left(\frac{\sqrt{2E_e}A_0}{\omega},-\frac{A_0^2}{4\omega}\right)\int e^{i\left[(E_e+E_\text{N}+A_0^2/4-E_v-n\omega)t\right]}\left[A(t)\sqrt{2E_e}+\frac{A^2(t)}{2}\right]dt\bigg\vert^2\label{eq:Fn}
\end{align}
\normalsize
with $J_n$ a generalized Bessel function~[36]. To get rid of the summation over nuclear and electron momenta in Eq.~\eqref{eq:SFACW}, we exploited the symmetry of the initial vibrational and electronic wave functions along with the following properties of the generalized Bessel function and the vector potential: $J_n(-u,v)=(-1)^nJ_n(u,v)$ and $A(t+\pi/\omega)=-A(t)$. We also exploited the relation $J_n(u,-v)=(-1)^nJ_{-n}(u,v)$ to change the sign of $n$ in the phase of the time integral.

In Eq.~\eqref{eq:Fn}, the time integral picks out the energy conservation lines as defined by 
\begin{equation}
E_e+E_\text{N}+U_p-E_v-n\omega=0, 
\end{equation}
where $n$ corresponds to $n$-photon absorption and $U_p=A_0^2/4$ represents the ponderomotive energy shift that the electron acquires due to the oscillating field. The generalized Bessel function accounts for the observed laser parameter dependence of the individual multiphoton peaks. We point out that for larger molecules, the Fourier transform in Eq.~\eqref{eq:SFACW} is replaced by the Fourier transform of the total wave function and that the sum $E_\text{N}+E_e$ in the phase of the time integral is replaced by the total energy. Additionally, the first argument of the generalized Bessel function and the laser-molecule interaction part, $V_\text{Ext}$, of the time integral will have additional terms for any particle that couples directly to the laser field. From these observations, it immediately follows that our findings generalize to larger molecules.


\appendix
